\documentclass[prb,nopacs,twocolumn,preprintnumbers,amsmath,amssymb]{revtex4}

\usepackage{epsfig}
\usepackage{graphicx}% Include figure files
\usepackage{dcolumn}% Align table columns on decimal point
\usepackage{color}
\usepackage{bm}% bold math
\usepackage{multirow}
\usepackage{amsmath}

\begin{document}

\title{Spin memory and spin-lattice relaxation in two-dimensional hexagonal crystals}

\author{H. Ochoa,$^1$ F. Guinea,$^1$ V. I. Fal'ko,$^{2}$}

{\affiliation{$^1$ Instituto de Ciencia de Materiales de Madrid. CSIC. Sor Juana In\'es de la Cruz 3. 28049 Madrid. Spain.\\
$^2$Physics Department, Lancaster University, Lancaster, LA1 4YB, UK.}

\begin{abstract}
We propose a theory of spin relaxation of electrons and holes in two-dimensional hexagonal crystals such as atomic layers of transition metal dichalcogenides (MoS$_2$, WSe$_2$, etc). We show that even in intrinsically defectless crystals, their flexural deformations are able to generate spin relaxation of carriers. Based on symmetry analysis, we formulate a generic model for spin-lattice coupling between electrons and flexural deformations, and use it to determine temperature and material-dependent spin lifetimes in atomic crystals in ambient conditions.
\end{abstract}
%\pacs{73.20.-r; 73.20.Hb; 73.23.-b; 73.43.-f}

\maketitle

Atomically thin crystals, derived by exfoliation from Van der Waals-coupled layered materials,\citep{Geim_Grigorieva_Nat2013} represent a very natural and truly two-dimensional systems for the implementation in electronic devices. Started with graphene,\citep{Novoselov2004,graphene_rev} the family of atomically thin 2D crystals already includes silicene,\citep{Vogt_etal,Fleurence_etal} graphane C$_2$H$_2$,\citep{Elias_etal_2009} gemanane Ge$_2$H$_2$,\citep{Bianco_etal} monolayers of hBN,\citep{Watanabe_etal} transition metal dichalcogenides MX$_2$ (M=Mo,W,Ta; X=S,Se,Te),\citep{Gordon_etal,Wang_etal_2012} and bilayers of gallium chalcogenides, Ga$_2$X$_2$.\citep{Zolyomi_etal} Inspite of different chemical composition, these crystals share common honeycomb-like lattice and several similar features in their electronic properties. The low-energy electron spectra near the bottom of conduction band or/and top of the valence band of these two-dimensional hexagonal crystals (2DHC) appear in the valleys near the corners $\pm$K of the hexagonal Brillouin zone (BZ), related by time-inversion, where electrons in each valley display the same quantum dynamics specific to systems with broken time-inversion symmetry.\cite{Iordanskii_Koshelev,Morpurgo_Guinea,McCann_etal_2006,McCann_etal_2007,Vozmediano_rep} For the electron orbital dynamics, such fictitious time-inversion symmetry breaking can be associated with the appearence of a valley-antisymmetric pseudomagnetic field in deformed graphene,\cite{Vozmediano_rep} or with trigonal warping of the electron dispersion.\citep{McCann_etal_2007,Kormanyos_etal_2013}

Also, a fictitious time-inversion symmetry breaking can be associated with the opposite site of the spin-orbital (SO) splitting for the electron states in $\pm$K valleys which is stronger in materials with heavier X and M elements.\citep{Zhu_etal_2011,Feng_etal_2012} Another specific feature of SO coupling in 2DHCs arises from their $z \to -z$ mirror reflection symmetry. Since the in-plane components $s_{x,y}$ of the electron spin invert their sign upon $z \to -z$  reflection, only $s_z$ appears in generic $\mathbf{k}\cdot\mathbf{p}$ Hamiltonians for electrons, 
\citep{Kane_Mele,Huertas-Hernando_etal_prb,Min_etal,Xiao_etal_2012,Ochoa_Roldan_2013}
\begin{align}
\mathcal{H}\left(\pm\mathbf{K}+\mathbf{p}\right)=\mathcal{H}_{band}\left(\pm\mathbf{K}+\mathbf{p}\right)
+\epsilon_z\left(\mathbf{p}\right)\hat{\mathcal{L}}_z\hat{s}_z,
\label{eq:Hamiltonian_int}
\end{align}
where $\mathbf{p}$ is the electron momentum counted from the centre of the valley. Here, the first term describes the electron/hole orbital motion in the band: $\mathcal{H}_{band} \approx \mathbf{p}^2/2m_*$ for gapful 2DHCs, in contrast to the linear dispersion of electrons in graphene and silicene, where $\mathcal{H}_{band} \approx v\mathbf{p} \mathbf{\sigma}$ with Pauli matrices $\sigma_{x,y,z}$ acting on the sublattice component of the electronic wave function. The second term in Eq.~\eqref{eq:Hamiltonian_int} takes into account atomic SO coupling, where the microscopic form of the angular momentum operator, $\hat{\mathbf{\mathcal{L}}}_z=\pm 1$ for $\pm$K in MX$_2$ and $\hat{\mathbf{\mathcal{L}}}_z=\pm\sigma_z$ in graphene and silicene, and size of the coupling constant, $\epsilon_z\left(\mathbf{p}\right)\approx \epsilon_z+p^2\delta m_*^{-1}$, depend on the material and band-specific orbital composition of the electron Bloch states.\citep{RMA13,Kormanyos_etal_2013,Cappelluti_etal_2013} In each of the two valleys $\pm$K, this produces spin splitting, $|\epsilon_{z}|$, for electrons (holes), separating their constant-energy contours in the momentum space by $\epsilon_{z}/v$, where $v(\varepsilon)$ is the electron velocity $\hat{\mathbf{v}}=\partial H_{band}/\partial\mathbf{p}$ at energy $\varepsilon$.

The Hamiltonian in Eq. (\ref{eq:Hamiltonian_int}) suggests that, for all 2DHCs, $s_z$ component of the electron spin conserves, resulting in a long spin memory.\citep{Mak_etal_2012,Zeng_etal_2012,Cao_etal_2012} This hints that 2DHCs offer a promising materials platform for spintronics applications. A long spin memory of photo-excited electrons and holes, at the time scale of several nanoseconds, has been already observed \cite{Mak_etal_2012} in MoS$_2$. At the same time, it has been noticed\citep{Falko,Ochoaetal2,Fratini_etal} that out-of-plane (flexural) deformations of graphene locally break its $z \to -z$ mirror symmetry and couple the in-plane $s_{x,y}$ spin components of the electron to its lateral orbital motion. A similar effect can generate spin relaxation of charge carriers in 2DHCs-based spintronic devices free of intrinsic structural defects. In this paper, we study how out-of-plane spin relaxation is generated by spin-lattice coupling with flexural phonons or substrate-induced bending of 2DHCs. This study is based on the symmetry analysis of possible spin-lattice couplings permissible in weakly perturbed 2DHCs, and it is used to identify relevant parametric regimes of spin relaxation of electrons and holes. 

\begin{figure}
\begin{centering}
\includegraphics[width=0.6\columnwidth]{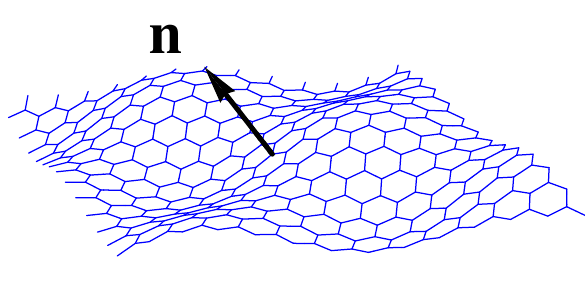}
\par\end{centering}
\caption{\label{fig:honeycomb}Geometry of local spin-coordinate system of electrons in wrinkled 2DHCs.}
\end{figure}

%\begin{center}
%\begin{table}
%\begin{tabular}{|c||c|c||c|c||c|c|}
%\hline
%Material&\multicolumn{2}{c||}{$m^{*}/m_0$}&\multicolumn{2}{c||}{$\epsilon_z$ }&\multicolumn{2}{c||}{$n_0$}\\
%\hline
%\hline
%MoS$_2$&0.48, {\green 0.48}&0.59&2.8&145.6, {\red 150}, {\blue 148}&0.56&35.92\\
%\hline
%WS$_2$&0.32&0.43&38.4&420.8, {\red 430}, {\blue 426}&5.14&75.66\\
%\hline
%MoSe$_2$& & & &{\red 180}, {\blue 183}& & \\
%\hline
%WSe$_2$& & & &{\red 460}, {\blue 456}& & \\
%\hline
%MoTe$_2$& & & & & & \\
%\hline
%WTe$_2$& & & & & & \\
%\hline
%\hline
%Bands&cb&vb&cb&vb&cb&vb\\
%\hline
%\end{tabular}
%\caption{\label{tab:parameters}Parameters from Viktor in black, taken from Xiao \textit{et al} in red,\citep{Xiao_etal_2012,Feng_etal_2012} from Zhu in blue,\citep{Zhu_etal_2011} from Kaasbjerg \textit{et al} in green.\citep{Kaasbjerg_etal_2012}}
%\end{table}
%\end{center}

{\em Spin-lattice coupling} between electrons and wrinkles can be incorporated in Eq.~\eqref{eq:Hamiltonian_int} by treating the 2DHC as a flexible membrane described by local vertical displacement $h(x,y)$ and local normal vector $\mathbf{n}=\left(-\partial_x h,-\partial_y h,1\right)$, see Fig.~\ref{fig:honeycomb}. Then, we use the global coordinate system for the 3D electron spin $\mathbf{s}$ to write down its coupling to the local angular momentum oriented along $\mathbf{n}(x,y)$,
\begin{eqnarray}
H_{SO}=\epsilon_z\hat{\mathcal{L}}_z\hat{\mathbf{s}}\cdot\mathbf{n}
\approx 
\epsilon_z\hat{\mathcal{L}}_z\hat{s}_z + \delta H_{g};
\label{eq:normalSO}
\\
\delta H_{g} = -\epsilon_z\hat{\mathcal{L}}_z\left(\partial_xh\hat{s}_x+\partial_yh\hat{s}_y\right). 
\nonumber
\end{eqnarray}
Here, the inhomogenous term $\delta H_{g}$ is responsible for spin-lattice relaxation, whereas the first term $\epsilon_z\hat{\mathcal{L}}_z  s_z$ sets a global quantization axis for the electron spin. Additionally, we take into account contributions towards spin-phonon coupling arising from the modification of the orbital composition of the Bloch states of electrons, due to the mixing of bands (and corresponding atomic orbitals) by mutual displacements of atoms in the 2DHC lattice. Such couplings depend on the local curvature tensor $h_{ij}^{''} \equiv \partial_i \partial_j h$ of the deformed 2DHC rather than $\mathbf{n}$,\citep{Ochoaetal2} since tilting of a crystal does not change orbital composition of electronic states. Phenomenologically, such additional couplings,
\begin{gather}
\label{eq:Hamiltonian_ext}
\delta H_{o}=\lambda_{\parallel}\left[2h_{xy}^{''}\hat{s}_x+\left(h_{yy}^{''}-h_{xx}^{''}\right)\hat{s}_y\right]\hat{\mathbf{\mathcal{L}}}_z 
\nonumber\\+\hbar\beta\left(\mathbf{v}\times\mathbf{s}\right)_z\nabla^2h \\
+\hbar\tilde{\beta}[\left(\hat{v}_x\hat{s}_y+\hat{v}_y\hat{s}_x\right)\left(h_{xx}^{''}-h_{yy}^{''}\right)
+\left(\hat{v}_y\hat{s}_y-\hat{v}_x\hat{s}_x\right)2h_{xy}^{''}];\nonumber
\end{gather}appear as invariants of the symmetry group of a 2DHC, built from the components of the electron spin operator $s_{x,y}$, velocity operator, and curvature tensor.
The corresponding 2DHC symmetry group includes lattice translations, amended by $C_{6v}$ rotations and reflections for graphene and by  $D_{3h}=D_3\times\sigma_h$ for MX$_2$ and Ga$_2$X$_2$. Table~\ref{tab:operators} lists irreducible representations (IrReps) of $D_{3h}$ (for $C_{6v}$ see Ref.~\onlinecite{Ochoaetal2}) and classify relevant operators with respect to their transformation properties, including $z \to -z$ mirror reflection and time inversion. Consequently, all terms in Eq.~(3) are scalar products of operators belonging to IrReps in Tab.~\ref{tab:operators}. Note that the terms in the first row of Eq.~(3) can be written as $\lambda_{\parallel}\hat{\mathbf{\mathcal{L}}}\cdot\hat{\mathbf{s}}$, where  $\hat{\mathbf{\mathcal{L}}}=\Lambda_z\left(2\partial_x\partial_yh,\partial_y^2h-\partial_x^2h\right)$ has the properties of an in-plane component of atomic angular momentum operator, and the other two terms are similar to Bychkov-Rashba\citep{Bychkov_Rashba,Rashba} and Dresselhaus\citep{Dresselhaus} SO coupling. 
\footnote{Note that there is another possible combination of velocity and spin operators which can be written as $\hat{\mathbf{v}}\cdot\hat{\mathbf{s}}$, but this is a complete pseudo-scalar which cannot be coupled with these excitations.} In those 2DHC bands where electrons originate from atomic P$^{x,y}$ and D$^{xy,x^2-y^2}$ orbitals (e.g., valence band in MoS$_2$), the influence of $\delta H_{o}$ should be less than that of $\delta H_{g}$. In the bands where electrons originate from S, P$^z$ or D$^{z^2}$ orbitals, SO coupling arises from their weak mixing with high energy orbitals, and hence $\delta H_{g}$ and $\delta H_{o}$ should be treated on equal footing.  

\begin{center}
\begin{table}
\begin{tabular}{|c|c|c||c|c|}
\hline
\multicolumn{1}{|c|}{Irrep}&\multicolumn{2}{c||}{$D_{3h}=D_3\times\sigma_h$}&Material&$\epsilon_z$ (meV)\\
\hline
\hline
$A_1'$&$\Lambda_z\hat{s}_z$&&\multirow{2}{*}{e-MoS$_2$}&\multirow{2}{*}{3}\\
\cline{1-3}
$A_2'$&&$\Lambda_z$, $\hat{s}_z$&&\\
\cline{1-5}
$A_1''$&$\hat{\mathbf{v}}\cdot\hat{\mathbf{s}}$&&\multirow{2}{*}{h-MoS$_2$}&\multirow{2}{*}{140}\\
\cline{1-3}
$A_2''$&$\nabla^2 h$, $\left(\hat{\mathbf{v}}\times\hat{\mathbf{s}}\right)_z$&&&\\
\cline{1-5}
$E'$&&$\left(\begin{array}{c}
\hat{v}_x\\
\hat{v}_y \end{array}\right)$&h-WS$_2$&430\\
\cline{1-5}
$E''$&$\left(\begin{array}{c}
\partial_x^2h-\partial_y^2h\\
2\partial_x\partial_y h\end{array}\right)$,
&$\left(\begin{array}{c}
\hat{s}_x\\
\hat{s}_y \end{array}\right)$&h-MoSe$_2$&180\\
\cline{4-5}
&$\left(\begin{array}{c}
\hat{v}_x\hat{s}_y+\hat{v}_y\hat{s}_x\\
\hat{v}_y\hat{s}_y-\hat{v}_x\hat{s}_x \end{array}\right)$&&h-WSe$_2$&460\\
\cline{1-5}
\multicolumn{1}{|c|}{$t\to-t$ }&\multicolumn{1}{c|}{even}&\multicolumn{1}{c||}{odd}\\
\cline{1-3}
\end{tabular}
\caption{Left: Classification of the electronic operators and flexural displacements according to their transformation properties under time reversal $t\rightarrow-t$ and symmetry operations of the 2DHCs without inversion symmetry. Operators belonging to representations $A_{1,2}'$ and $E'$ are $z\to-z$ even, and those belonging to $A_{1,2}''$ and $E''$ are odd. Right: SO splitting in the conduction (e) and valence (h) bands in 2DCHs of MX$_2$.\citep{Kormanyos_etal_2013,Zhu_etal_2011,Feng_etal_2012}}
\label{tab:operators}
\end{table}
\end{center}

As they stand in Eqs. (\ref{eq:normalSO},\ref{eq:Hamiltonian_ext}), the spin-lattice coupling terms $\delta H_{g,o}$ can be used to evaluate the rate of spin relaxation of electrons due to the short-wavelength ripples with a Fourier spectrum $h_q$ in the range of wave numbers $q\gg\epsilon_{z}/v$. To describe spin of electrons flying accross such short-wavelength ripples, we use a spin-coordinate frame related to the median orientation of the 2DHC, averaged over many ripples periods.  
In contrast, it is more practical to analyse the influence of long-wavelength wrinkles, with  $q<\epsilon_{x}/v$, in the local, adiabatically varying spin frame, adjusted to the local flake orientation. The electron spinor states in the global and local frames are related by the non-Abelian gauge transformation
$\hat{U}=e^{\frac{i}{2}\mathbf{\nu}\cdot\hat{\mathbf{s}}}$, $\mathbf{\nu}=\left(\partial_yh,-\partial_x h\right)$,
which diagonalises Eq.~\eqref{eq:normalSO} into $H_{SO}=\epsilon_z\hat{\mathcal{L}}_z\hat{s}_z$, but also produces an additional smaller perturbation,
\begin{gather}
\hat{U}\mathcal{H}_{band}\left(\pm\mathbf{K}+\mathbf{p}\right)\hat{U}^{\dagger}\approx
\mathcal{H}_{band}\left(\pm\mathbf{K}+\mathbf{p}\right)+\delta \tilde{H}_{g} \nonumber\\
\delta \tilde{H}_{g} = \frac{1}{2}\left\{\frac{\partial \mathcal{H}_{band}}{\partial \mathbf{p}},\hat{U}\left(-i\hbar \mathbf{\partial}\right)\hat{U}^{\dagger}\right\}\\
=\frac{\hbar}{2}\left[\hat{v}_y\hat{s}_x\partial_y^2h-\hat{v}_x\hat{s}_y\partial_x^2h+\left(\hat{v}_x\hat{s}_x-
\hat{v}_y\hat{s}_y\right)\partial_x\partial_y h\right].\nonumber
\label{eq:Hamiltonian_g}
\end{gather}
Note that, upon gauge transformation $\hat{U}$, spin-lattice coupling $\delta H_{o}$ remains almost unchanged (only terms in the higher order in $qh_q$ can appear), hence $\delta \tilde{H}_{g}$ is the leading term in the Taylor expansion of $\hat{U}\mathcal{H}_{band}\hat{U}^{\dagger}-\mathcal{H}_{band}$ in small $qh_q$. Also, $\delta \tilde{H}_{g}$ has the same structure as a combination of Bychkov-Rashba and Dresselhaus terms in the phenomenological Eq.~(3), with the universal coupling constants, $\beta=\tilde{\beta}=-1/4$, which manifests the geometrical origin of this coupling. Therefore, in the following analysis of spin relaxation induced by smooth ripples, we combine $\delta \tilde{H}_{g}$ and $\delta H_o$ by redefining the coupling constants $\beta$ and $\tilde{\beta}$.

{\em Spin-lattice relaxation} of electrons is determined by the cumulative contribution of both short- and long-wavelength lattice deformations, which produces the sum, $\tau_s^{-1} = \tau_{d}^{-1}  +  \tau_{b}^{-1}$, of 'diffusive' spin relaxation assisted by external charge disorder in the substrate and 'ballistic' contribution determined by a simultaneous momentum $|\mathbf{p}-\mathbf{p}'| \sim q$ and spin transfer to the ripples.

\begin{figure}
\begin{centering}
\includegraphics[width=1\columnwidth]{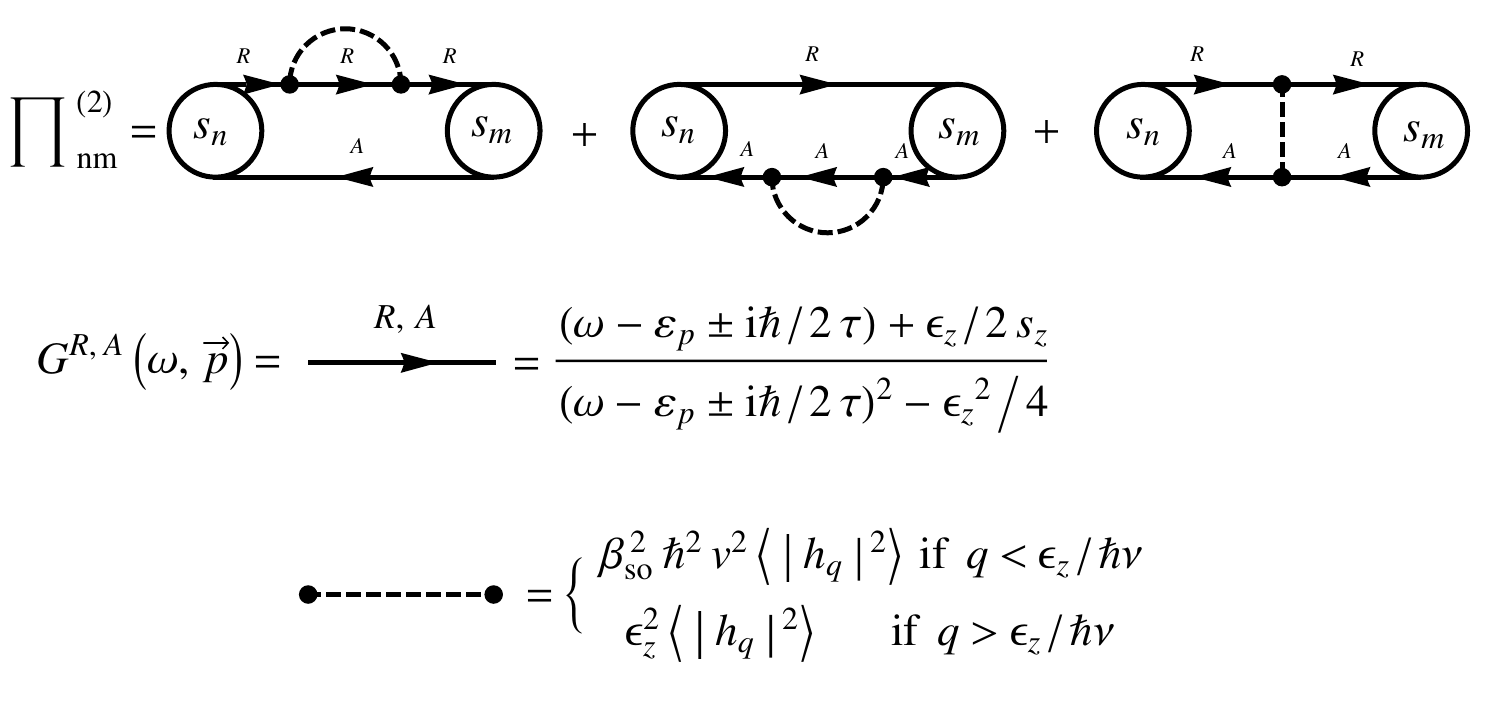}
\par\end{centering}
\caption{\label{fig:diagrams} The three diagrams which contribute to $\Pi$ operator to the lowest order in the spin-lattice coupling.}
\end{figure}

{\em The diffusive contribution} to the spin relaxation rate can be estimated using the framework of diagrammatic perturbation theory applied to the analysis of a disorder averaged spin-density matrix of electrons, $\frac{1}{2} \vec{\rho} \cdot \vec{\sigma}$.  Diagramatically, Fig.~\ref{fig:diagrams}, spin-lattice coupling is incorporated in the polarisation operator $\Pi$, which governs spin diffusion, $[\partial_t - \Pi]\vec{\rho}(t)= \vec{\rho}(0) \delta(t)$. Without spin-lattice coupling, $\Pi \approx D\nabla^2$, where $D=\frac12 \ell v$ and $\ell$ stands for the electron diffusion coefficient and mean free path, respectively. Valley-dependent spin splitting, $\epsilon_z \Lambda_z s_z$, generates independent precession of $\vec{\rho}_{\pm}$  in $\pm$K valleys. Spin-lattice relaxation of electrons, assisted by disorder, is incorporated into $\Pi$ via three diagrams, Fig.~\ref{fig:diagrams}, where solid lines indicate the free electron Green functions, the wavy lines are correlators $\langle h({\mathbf{r}})h({\mathbf{r'}}) \rangle$ and dots are spin-lattice coupling vertices cosrresponding to Eqs.~\eqref{eq:normalSO}-(3).  The kinetics of spin polarization of carriers is, then, described by 
\begin{equation}
\left[ \partial_t - D\nabla^2 \right]\vec{\rho}_{\pm} \pm\lambda_{Z_2} \mathbf{n}_z \times\vec{\rho} _{\pm}
+ \tau_{d}^{-1}\vec{\rho} =\vec{\rho}_{\pm}(0) \delta(t),
\end{equation}
where $\pm$ identifies $\pm$K valley, and we have neglected the difference between the in-plane and out-of-plane spin relaxation rates, regarding the fact that, in all possible regimes, it should be superceeded by a faster spin precession due to $\epsilon_z$ term in Eq.~\eqref{eq:Hamiltonian_int}. The three diagrams in Fig.~\ref{fig:diagrams} lead to
\begin{gather}
\tau_{d}^{-1}=\sum_{\mathbf{q}}\mathcal{M}\left(q\right)\frac{\tau q^2}{N}\left\langle\left|h_{\mathbf{q}}\right|^2\right\rangle\times\begin{cases}
q^2\beta_{SO}^2v^2&q<\frac{\epsilon_z}{\hbar v}\\
\epsilon_z^2/\hbar^2&q>\frac{\epsilon_z}{\hbar v}
\end{cases}
\nonumber\\
\mathcal{M}\left(q\right)=\frac{1+\ell^2q^2+\frac{\tau^2\epsilon_z^2}{\hbar^2}}{1+2\left(\ell^2q^2+\frac{\tau^2\epsilon_z^2}{\hbar^2}\right)+\left(\ell^2q^2-
\frac{\tau^2\epsilon_z^2}{\hbar^2}\right)^2}
\nonumber\\
\beta_{SO}^2\equiv\beta^2+\tilde{\beta}^2+\frac{\lambda_{\parallel}^2}{\hbar^2v^2}
\label{eq:tau_a}
\end{gather}
The above expressions link together D'yakonov-Perel'\citep{DyakonovPerel,Dyakonov} and Elliot-Yafet\citep{Elliot,Yafet} regimes discussed in the theory of spin relaxation in disordered systems. Indeed, for $q\ell\ll1$ and $\tau\epsilon_z\ll\hbar$, spin relaxation takes place over several momentum scattering events while electron diffuses in an interval of space with almost homogeneous SO coupling which causes its spin to precess randomly, so spin relaxation obeys the D'yakonov-Perel' relation:\citep{DyakonovPerel,Dyakonov} $\tau_{d}^{-1}\propto\tau$. On the contrary, for $q\ell> 1$ or $\tau\epsilon_z>\hbar$, spin flips take place over single scattering events involving external disorder, and  spin relaxation obeys the Elliot-Yaffet relation:\citep{Elliot,Yafet} $\tau_{d}^{-1}\propto\tau^{-1}$.

{\em The ballistic contribution} is determined by a simultaneous momentum, $|\mathbf{p}-\mathbf{p}'| \sim q$, and spin transfer to the lattice upon electron scattering off the ripples,
\begin{align}
\tau_{b}^{-1}=\frac{2\pi}{N\hbar}\sum_{\mathbf{q}} e^{-\frac{1}{q\ell}}
\left|\langle\mathbf{p}+\mathbf{q}  \uparrow \left| \delta H_{g}\right| \mathbf{p} \downarrow  \rangle \right|^2
\delta\left(\varepsilon_{\mathbf{p}+\mathbf{q}}^{\uparrow}-\varepsilon_{\mathbf{p}}^{\downarrow}\right).
\nonumber
\end{align}
Here, a factor $e^{-\frac{1}{q\ell}}$ takes into account the fact that this contribution does not involve any externally promoted momentum transfer.
This contribution is generated by the short-wavelegth flexural deformations $h_q$ with $q>\epsilon_{z}/\hbar v$, which permit electron's scattering between isoenergy lines separated by $\epsilon_z/\hbar v$ on the momentum plane near the $\pm$K points (the local spin quantization axis is defined as a normal to the median plane averaging  2DHC wrinkles over $\hbar v/\epsilon_z$ scale).

\begin{figure}
\begin{centering}
\includegraphics[width=0.9\columnwidth]{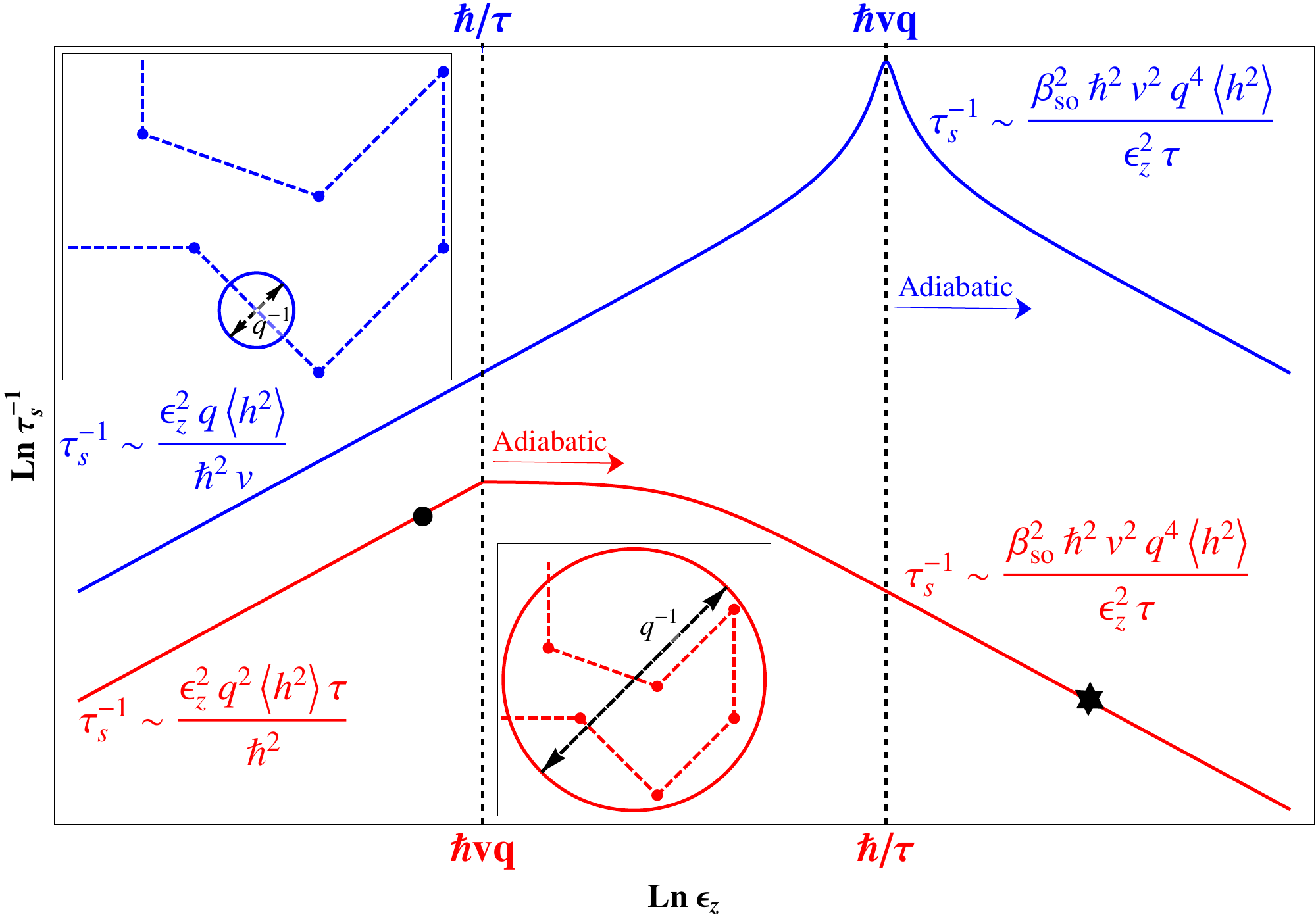}
\includegraphics[width=0.9\columnwidth]{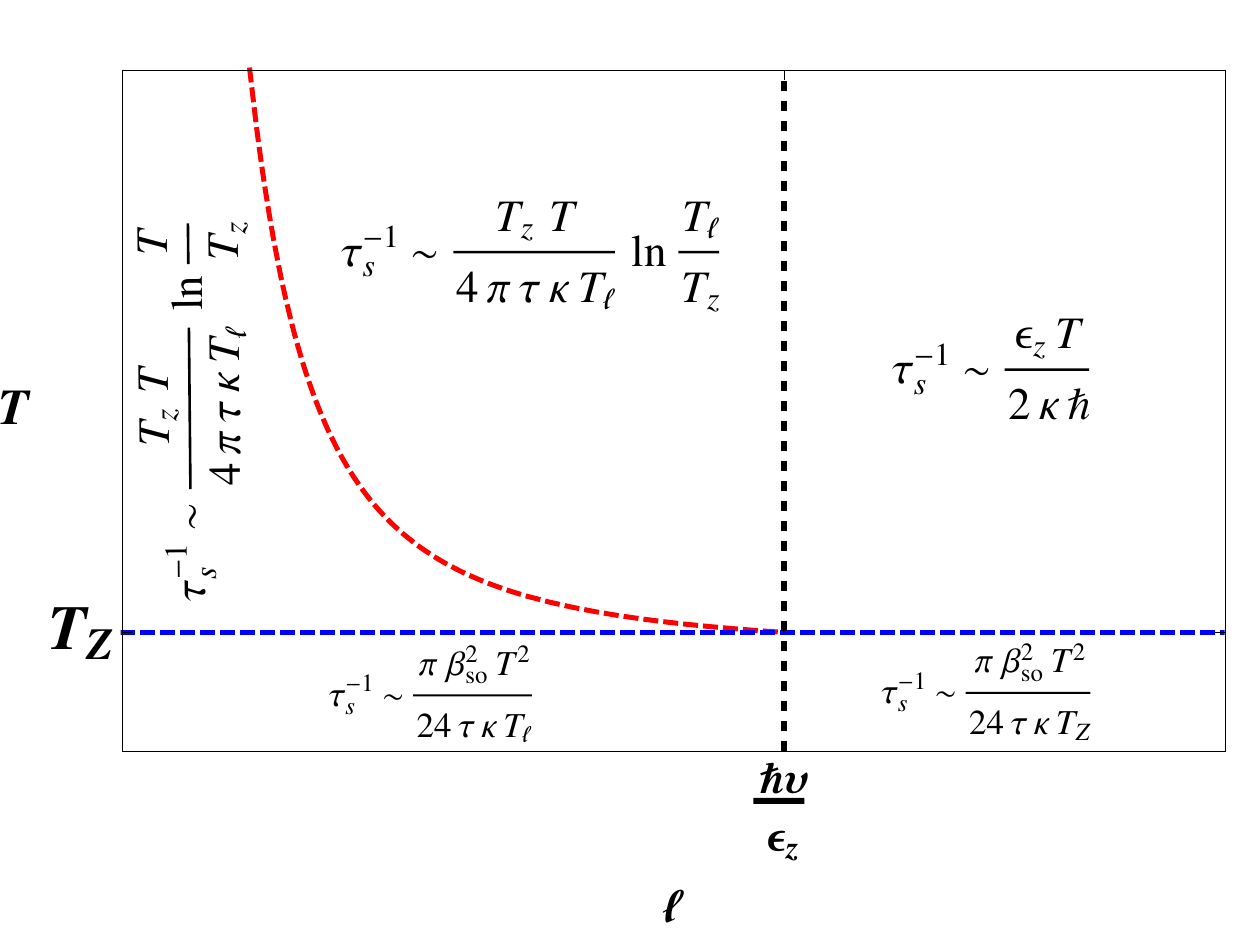}
\par\end{centering}
\caption{\label{fig:wrinkles} Top: Schematic behavior of spin relaxation induced by wrinkles of typical size $q^{-1}$ and height $\sqrt{\left\langle h^2\right\rangle}$. The top and bottom lines correspond to $q\ell<1$ and $q\ell>1$, respectively. The experimental situation\citep{Brivio_etal_2011,Radisavljevic_etal_2011,Zhang_etal,Fuhrer} for electrons and holes in MoS$_2$ is denoted by a dot and a star, respectively. Bottom: Spin relaxation induced by flexural phonons for different regimes of temperature and disorder. Dashed line represents $T_{\ell}$.}
\end{figure}

To analyse spin relaxation of electrons produced by static wrinkles in 2DHC we distinguish two asymptotic regimes characterized by the ratio $q\ell$, see Fig.~\ref{fig:wrinkles}. In the case of $q\ell>1$ and $\epsilon_z<\hbar v/q$, $\tau_b^{-1}\gg\tau_d^{-1}$, whereas for $q\ell\ll1$, $\tau_b^{-1}$ is exponentially suppressed.
Moreover, in both cases we find a non-monotonic dependence of $\tau_s^{-1}$ on $\epsilon_z$, due to a double role it plays: on the one hand $\epsilon_z$ determines the strength of the spin-lattice coupling according to Eq.~\eqref{eq:normalSO}, and on the other it represents the intrinsic SO splitting leading to a pseudo-Zeeman field which protects spin polarization.

{\em Spin relaxation due to flexural phonons} is evaluated taking into account that their quadratic dispersion, $\omega_{\mathbf{q}}=\sqrt{\frac{\kappa}{\rho}}\left|\mathbf{q}\right|^2$ (where $\kappa$ is the bending rigidity of the system and $\rho=M/A_c$ the mass density) and the resulting low frequencies, allow us to treat them as quasi-static deformations parametrised by spectral density $\langle \left| h_{\mathbf{q}}\right|^2 \rangle$,
\begin{align*}
\left\langle\left|h_{\mathbf{q}}\right|^2\right\rangle=\frac{\hbar}{2M\omega_{\mathbf{q}}}\left[2n_B\left(\frac{\omega_{\mathbf{q}}}{T}\right)+1\right],
\end{align*}
where $n_B$ is the Bose-Einstein distribution function. For the sake of convenience, we introduce two characteristic temperature scales:\footnote{Boltzmann constant K$_B$ is set to 1.} $T_{\ell}\equiv\ell^{-2}\hbar\sqrt{\kappa/\rho}$ and $T_Z\equiv\sqrt{\kappa/\rho}\epsilon_z^2/\hbar v^2$. For $T\ll T_Z$, spin relaxation is dominated by long wavelength ($q<\epsilon_z/\hbar v$) phonons, whereas at $T\gg T_Z$, short wavelength modes dominate. The two regimes of disorder are then defined by the ratio $T_{Z}/T_{\ell}$.
Overall, after the integration over thermally excited flexural phonon modes, we find\begin{align}
\tau_{d}^{-1}\approx\begin{cases}
\frac{\pi\beta_{SO}^2T^2}{24\tau\kappa T_{\ell}}&T\ll T_{Z}\ll T_{\ell};\\
\frac{T_ZT}{4\pi\tau\kappa T_{\ell}}\times\ln\left(\frac{T}{T_Z}\right)&T_{Z}\ll T\ll T_{\ell};\\
\frac{T_ZT}{4\pi\tau\kappa T_{\ell}}\times\ln\left(\frac{T_{\ell}}{T_Z}\right)&T_Z\ll T_{\ell}\ll T;\\
\frac{\pi\beta_{SO}^2T^2}{24\tau\kappa T_Z}&T_{\ell},T\ll T_Z;\\
\frac{K_BT}{4\pi\kappa\tau}&T_{\ell}\ll T_Z\ll T.
\end{cases}
\label{eq:diff}
\end{align}

%\begin{figure}
%\begin{centering}
%\includegraphics[width=0.9\columnwidth]{phonons}
%\includegraphics[width=1\columnwidth]{ballistic}
%\par\end{centering}
%\caption{\label{fig:diffusive} Spin relaxation induced by flexural phonons for different regimes of temperature and disorder. The dashed red line represents $T_{\ell}$.}
%\end{figure} 

Among all these regimes, only the asymptotic of $T\gg T_{Z}$ and $\hbar v/\epsilon_z>\ell$ is dominated by ballistics,
\begin{gather}
\label{eq:bal}
\tau_{b}^{-1}=\frac{2\varepsilon T_{Z}}{\hbar\kappa}\cdot f\left(\frac{T_{M}}{T},\frac{T_{m}}{T},\frac{T_{\ell}}{T}\right),\\
f\left(X,x,z\right)\equiv\int_{x}^{X}\frac{dy}{2\pi}\frac{\frac{e^y+1}{e^y-1}\times e^{-\sqrt{z/y}}}{\sqrt{\left(y-x\right)\left(X-y\right)}},\nonumber\\
T_{M,m}=\frac{4\varepsilon T_{Z}}{\epsilon_z}\left(\sqrt{\varepsilon/\epsilon_{z}}\pm\sqrt{\varepsilon/\epsilon_{z}-1}\right)^2.
\nonumber
\end{gather}
Having compared the diffusive and ballistic contributions\footnote{When $z\ll X,x$, $f\approx
\sqrt{\frac{1}{Xx}}$ for $X,x\ll 1$, and $f\approx\sqrt{\frac{1}{\pi X}}e^{-x}\mbox{Erf}\left(\sqrt{X-x}\right)$ for $X,x\gg 1$. If $z\gg X,x$ then $f$ is exponentially supressed. For $T> T_M> T_{\ell}$, the only regime when $\tau_b^{-1}$ is not exponentially small, we find that
$\frac{\tau_b^{-1}}{\tau_{d}^{-1}}\sim2\pi\frac{\tau\epsilon_z}{\hbar}\gg1$ since $T_{Z}\gg T_{\ell}$.}
in Eqs.~\eqref{eq:diff} and \eqref{eq:bal} and combined them in $\tau_{s}^{-1}=\tau_{d}^{-1}+\tau_{b}^{-1}$, we summarize the resulting behaviors in Fig.~\ref{fig:wrinkles}.

{\em Discussion.} In currently available MoS$_2$-based devices, the mobilities extracted from transport experiments\citep{Radisavljevic_etal_2011,Zhang_etal,Fuhrer} indicate that it is the diffusive contribution that limits spin lifetimes of electrons and holes in this material. For wrinkles with a typical height of 1 nm and lateral length scale of 10 nm, as reported in Ref.~\onlinecite{Brivio_etal_2011} and SO splitting quoted in Tab.~\ref{tab:operators}, we find\footnote{For electrons we employ the formula $\tau_s^{-1}\approx\frac{\epsilon_z^2\left\langle h^2\right\rangle\tau}{\mathcal{L}^2\hbar^2}$, whereas for holes $\tau_s^{-1}\approx\frac{\beta_{SO}^2\hbar^2v^2\left\langle h^2\right\rangle}{\epsilon_z^2\mathcal{L}^4\tau}$, with $\beta_{SO}^2=\frac{1}{16}$ (note that in addition the spin-phonon coupling Eq.~(3) should be included in $\beta_{SO}^2$).} $\tau_s\sim$ 1 ns. For the same conditions we estimate the spin lifetime of the holes as $\tau_s\geq10$ ns.
For perfectly flat MoS$_2$, flexural vibrations thermally activated at room temperature would also produce spin relaxation,\footnote{L. D. Landau and E. M. Lifshitz, Theory of Elasticity
(Pergamon Press, Oxford, 1959): we estimated bending rigidity $\kappa=\frac{Y\delta^3}{24\left(1-\sigma^2\right)}\approx27$ eV by describing MoS$_2$ crystal as a plate of thickness $\delta\approx6.75$ \AA,\citep{Benamen_etal_2011}, where $Y=0.33$ TPa  is the Young modulus\citep{Castellanos_etal_2012} and $\sigma=0.125$ is the Poisson ratio.\citep{Lovell_etal} For electrons this gives $T_{\ell}\sim 10$ K and $T_Z\sim10^{-3}T_{\ell}$, so we employ the formula $\tau_s^{-1}\approx\frac{T_ZT}{4\pi\tau\kappa T_{\ell}}\ln\left(\frac{T_{\ell}}{T_Z}\right)$. For holes, since $T_{Z}\sim 3000$ K, we employ the formula $\tau_s^{-1}\approx\frac{\pi\beta_{SO}^2T^2}{24\tau\kappa T_{Z}}$.} limiting spin lifetime of electron by $\tau_s\sim 5$ ns and for holes by $\tau_s\sim20$ ns. The latter estimate is produced without taking into account that an atomically flat substrate will quench bending modes\citep{Bruno} so that we expect 2DHCs-hBN structures to exhibit longer spin memory of charge carriers and therefore offer a perfect platform for spintronics devices.

{\em Acknowlegdments.} We thank T. Heinz, A. Morpurgo, R. Rold\'an, X. Xu and V. Zolyomi for useful discussions. This study was funded by JAE-Pre grant (CSIC, Spain), MINECO, Spain, through grant FIS2011-23713, ERC Advanced Grants 290846 and {\it Graphene and Beyond}, ERC Synergy Grant {\it Hetero2D}, and the European Graphene Flagship Project.

\bibliographystyle{apsrev}
\bibliography{biblio.bib}
\end{document}